\documentclass[aps,prd,amsmath,amssymb,showpacs,10pt,superscriptaddress]{revtex4-2}
\usepackage{graphicx}
\usepackage{color}
\usepackage{slashed}
\usepackage{subfigure}
\usepackage{enumerate}
\usepackage{physics}
\usepackage[font=footnotesize]{caption}
\usepackage[breaklinks,colorlinks,urlcolor=blue,linkcolor=red,anchorcolor=magenta,citecolor=blue]{hyperref}
\usepackage{cleveref}
\usepackage{orcidlink}

\crefname{table}{Table}{Tables}
\crefname{equation}{Eq.}{Eqs.}
\crefname{figure}{Fig.}{Figs.}
\crefname{section}{Sec.}{Secs.}
\allowdisplaybreaks
\begin{document}
\title{Study of the $e^{+}e^{-}\to J/\psi\,\pi^{+}\pi^{-}$ lineshape near the $D^{*}\bar{D}+c.c.$ threshold and possible signals for exotic hidden charm states}
\author{Jun Wang\orcidlink{0000-0002-0508-3349}}\email{junwang@ihep.ac.cn}
\affiliation{Institute of High Energy Physics, Chinese Academy of Sciences, Beijing 100049, People's Republic of China}
\affiliation{University of Chinese Academy of Sciences, Beijing 100049, People's Republic of China}
\author{Qiang Zhao\orcidlink{0000-0001-8334-2580}}\email{zhaoq@ihep.ac.cn}
\affiliation{Institute of High Energy Physics, Chinese Academy of Sciences, Beijing 100049, People's Republic of China}
\affiliation{University of Chinese Academy of Sciences, Beijing 100049, People's Republic of China}
\affiliation{Center for High Energy Physics, Henan Academy of Sciences, Zhengzhou 450046, People's Republic of China}

\begin{abstract}
  We investigate the lineshape of the $e^{+}e^{-}\to J/\psi\,\pi^{+}\pi^{-}$ cross section in the vicinity of the $D^{*}\bar{D}+c.c.$ threshold, where the ``so-called'' $G(3900)$ is observed in the $e^+e^-\to D\bar{D}$ channel. To take into account the possible $D^{*}\bar{D}+c.c.$ open channel effects or possible contributions from $G(3900)$, we include the intermediate meson loop transitions in $e^{+}e^{-}\to J/\psi\,\pi^{+}\pi^{-}$. As a consequence, a triangle singularity (TS) is fulfilled which can produce nontrivial structures in the $J/\psi\pi$ invariant-mass spectrum. Moreover, the TS transition also allows access to exotic quantum number of $(I,J^{P(C)})=(1,1^{-(-)})$ in the $J/\psi\pi$ invariant-mass spectrum. We present predictions for the $J/\psi\,\pi$ invariant-mass spectrum and our results clarify the different manifestations of the kinematic effects and genuine resonances. In particular, we show that resonance structures arising from the $P$-wave $D\bar{D}$ scatterings or hidden charm tetraquark state with $(I,J^{P(C)})=(1,1^{-(-)})$ can be identified by the $J/\psi\pi$ invariant mass spectrum. It can provide a theoretical guidance for future experimental search for these exotic candidates.
\end{abstract}

\maketitle

\section{Introduction}
The process of $e^+e^-$ annihilation provides a direct access to the production of the flavor neutral vector mesons in experiment. These vector mesons, which can directly couple to a virtual photon via the ``so-called" vector meson dominance (VMD), can thus be probed in their decays into various final states. While there have been tremendous achievements made in $e^+e^-$ annihilation experiments, there are also unanswered questions left behind. 
The exclusive channel of $e^{+}e^{-}\to J/\psi\pi^{+}\pi^{-}$ can serve as a unique laboratory for exploring exotic hadronic states beyond the conventional quark model, such as tetraquarks and hadronic molecules (see e.g. reviews on the relevant topics~\cite{Chen:2022asf,Liu:2019zoy,
Guo:2017jvc}).

The abnormal structure of $\psi(4230)$, which was historically denoted as $Y(4260)$, was first observed in $e^+e^-\to J/\psi\pi^{+}\pi^{-}$ by the BaBar Collaboration~\cite{BaBar:2005hhc}. With higher statistics its decay into $Z_c(3900)\pi\to J/\psi\pi\pi$ was found~\cite{BESIII:2013ris,BESIII:2013mhi,Belle:2013yex,BESIII:2013qmu,BESIII:2022qal,BESIII:2017bua}, which has motivated extensive theoretical studies. Concerning the nature of $\psi(4230)$ representative interpretations include a compact tetraquark proposed in Ref.~\cite{Maiani:2005pe}, with the corresponding $P$-wave spectrum further developed in Ref.~\cite{Maiani:2014aja}; a charmonium hybrid~\cite{Close:2005iz}, for which a quenched-lattice study identified a hybrid-like vector state at $4.33(2)\ \mathrm{GeV}$ and discussed its possible connection to $Y(4260)$~\cite{Chen:2016ejo}; hadro-charmonium, introduced in Ref.~\cite{Dubynskiy:2008mq} and subsequently applied to the mixed $Y(4260)$--$Y(4360)$ system~\cite{Li:2013ssa}; and a $D_1(2420)\bar D+c.c.$ hadronic molecule~\cite{Wang:2013cya,Cleven:2013mka}. A recent concise review is given in Ref.~\cite{Wang:2025clb}. In the molecular picture, the decay of the $D_1\bar D$ component naturally produces low-momentum $D\bar D^*+c.c.$ pairs and a cusp at the $D\bar D^*$ threshold~\cite{Wang:2013cya,Cleven:2013mka}. A recent global analysis of eight final states found that, within a framework including the $D_1\bar D$ threshold effect and interference with $\psi(4160)$, a single pole compatible with a $D_1\bar D$ molecular structure can describe the data in the $4.2\text{--}4.35\ \mathrm{GeV}$ region~\cite{vonDetten:2024eie}.

While most investigations have focused on the energy region around the $Y(4230)$, the center-of-mass (c.m.) energy range $\sqrt{s}=3.8\text{--}4.0\ \mathrm{GeV}$ remains less explored, especially in terms of high-statistics measurements. This kinematic window is of particular interest as it is close to the $D^{*}\bar{D}$ threshold. It was investigated that the $P$-wave threshold can produce enhancement, which is known as $G(3900)$ in the cross section lineshape of $e^+e^-\to D\bar{D}$~\cite{Zhang:2010zv,Zhang:2009gy}. Recently, with the high-statistics data for $e^+e^-\to D\bar{D}$ from BESIII~\cite{BESIII:2024ths} it is also proposed that $G(3900)$ may be interpreted as a $P$-wave hadronic-molecule candidate~\cite{Lin:2024qcq}. BESIII measured the $e^{+}e^{-}\to J / \psi \pi^{+}\pi^{-}$ cross section at the center-of-mass (c.m.) energies from 3.773 to 4.698 $\mathrm{GeV}$~\cite{BESIII:2022qal,BESIII:2016bnd}. In their fit a broad resonant structure near $3.9\ \mathrm{GeV}$ was introduced to improve the fit quality. However, the data point at $\sqrt{s}=3.8713\ \mathrm{GeV}$ can hardly be accommodated by the fitting and was excluded from the rest data points. Consequently, it remains uncertain whether the observed enhancement corresponds to a genuine resonance and whether it can be identified as the same state of $G(3900)$. Further experimental and theoretical investigations are required to resolve its nature.

Meanwhile, open-charm effects have a broad range of effects in $e^+e^-$ annihilation. They can not only play an important role in our understanding of the inclusive processes of $e^{+}e^{-}\to\text{hadrons}$ \cite{Peng:2024xui,Wang:2012mf,Wang:2011yh,Li:2007au}, but also affect various exclusive processes, among which those induced by the $D^{*}\bar{D}+c.c.$ rescatterings are of great interests~\cite{Li:2013zcr,Wang:2011yh,Wang:2012mf,Zhang:2010zv,Zhang:2009gy}.
While most of the previous studies have paid more attention to the two-body scatterings, their effects in the three-body final states remain to be explored~\cite{Wang:2025zbv}. One important feature arising from the three-body kinematics is the triangle singularity (TS)~\cite{Landau:1959fi,Cutkosky:1960sp,Coleman:1965xm,Wu:2011yx,Wang:2013cya,Wang:2013hga,Liu:2017rhx,Guo:2019twa,Du:2019idk,Dong:2020hxe}, which allows all the internal particles in a triangle loop to approach their on-shell condition simultaneously. The triangle singularities may significantly distort the lineshape of the cross section and create resonance-like structures in the two-body invariant mass spectrum \cite{Liu:2019dqc,Liu:2020orv,Liu:2016xly,Wu:2011yx,Wang:2013cya,Liu:2014spa,Liu:2013vfa,Guo:2014iya,Mikhasenko:2015oxp,Szczepaniak:2015eza,Guo:2015umn,Liu:2015taa,Liu:2015fea,Achasov:2015uua,Guo:2016bkl,Aceti:2012dj,Aceti:2016yeb,Bayar:2016ftu,Roca:2017bvy,Liu:2016dli,Liu:2015cah,Cao:2017lui,Dong:2020hxe,Nakamura:2021qvy}. Thus, it is essential to carefully disentangle genuine resonance contributions from the kinematic effects. In the process of $e^{+}e^{-}\to J / \psi \pi^{+}\pi^{-}$ the triangle process can have access to the exotic quantum number $(I,J^{P(C)})=(1,1^{-(-)})$ with hidden charm through the $P$-wave $D\bar{D}$ interaction. Since theoretical study of such exotic objects as either hadronic molecule or tetraquark state is strongly model-dependent, we are motivated to investigate the observables which are sensitive to its existence. We will show later that the triangle process is ideal for the search of the state of  $(I,J^{P(C)})=(1,1^{-(-)})$ which can be either $P$-wave hadronic molecule or tetraquark state. Furthermore, the invariant mass spectrum of $J/\psi\pi$ can distinguish the different effects produced by either the TS mechanism and a genuine state.

As follows, we first introduce our formalism in \cref{sec:formalism}. The numerical results and discussions are presented in \cref{sec:results}. A brief summary and conclusion are given in \cref{sec:summary}.

\section{Formalism}\label{sec:formalism}
\subsection{Effective Lagrangians}
We collect the relevant Lagrangians for the charmed meson and light pseudoscalar/vector meson couplings as follows~\cite{Cheng:2004ru,Zhang:2009gy,Zhang:2010zv,Wang:2011yh,Wang:2012mf}:
\begin{equation}
  \begin{aligned}
    {\cal L} =& - ig_{\mathcal{D}^*\mathcal{D}\mathcal{P}}(\mathcal{D}^i\partial^\mu \mathcal{P}_{ij}
    \mathcal{D}_\mu^{*j\dagger}-\mathcal{D}_\mu^{*i}\partial^\mu \mathcal{P}_{ij}\mathcal{D}^{j\dagger})
    +{\frac{1}{2}}g_{\mathcal{D}^*\mathcal{D}^*\mathcal{P}}
    \varepsilon_{\mu\nu\alpha\beta}\,\mathcal{D}_i^{*\mu}\partial^\nu \mathcal{P}^{ij}
    {\overleftrightarrow \partial}{}^{\!\alpha} \mathcal{D}^{*\beta\dagger}_j \\
    -& ig_{\mathcal{DDV}} \mathcal{D}_i^\dagger {\overleftrightarrow \partial}_{\!\mu} \mathcal{D}^j(\mathcal{V}^\mu)^i_j
    -2f_{\mathcal{D^*DV}} \epsilon_{\mu\nu\alpha\beta}
    (\partial^\mu \mathcal{V}^\nu)^i_j
    (\mathcal{D}_i^\dagger{\overleftrightarrow \partial}{}^{\!\alpha} \mathcal{D}^{*\beta j}-\mathcal{D}_i^{*\beta\dagger}{\overleftrightarrow \partial}{}{\!^\alpha} \mathcal{D}^j)
    \\
    +& ig_{\mathcal{D^*D^*V}} \mathcal{D}^{*\nu\dagger}_i {\overleftrightarrow \partial}_{\!\mu} \mathcal{D}^{*j}_\nu(\mathcal{V}^\mu)^i_j
    +4if_{\mathcal{D^*D^*V}} \mathcal{D}^{*\dagger}_{i\mu}(\partial^\mu \mathcal{V}^\nu-\partial^\nu
    \mathcal{V}^\mu)^i_j \mathcal{D}^{*j}_\nu \ ,
  \end{aligned}
\end{equation}
and the Lagrangian for the vector charmonium coupling to the charmed pseudoscalar and vector mesons reads:
\begin{equation}
  \mathcal{L}_{\psi}=g _{\psi \mathcal{D}^{*}\mathcal{D}}\varepsilon^{\mu \nu \alpha \beta}\partial_\mu \psi_{\nu}(\mathcal{D}^{*}_{\alpha}\overleftrightarrow{\partial}_{\beta}\mathcal{D}^{\dagger}+\mathcal{D} \overleftrightarrow{\partial}_{\alpha}\mathcal{D}^{*\dagger}_{\beta})\,,
\end{equation}
where $\mathcal{D}$ and $\mathcal{D}^{*}$ denote the pseudoscalar and vector charm meson fields, respectively, i.e.,
\begin{equation}
  \mathcal{D}=(D^{0},D^{+},D_{s}^{+}),\quad \mathcal{D}^{*}=(D^{*0},D^{*+},D_{s}^{*+}),
\end{equation}
$\mathcal{P}$ and $\mathcal{V}$ are $3\times 3$ matrices representing the light pseudoscalar octet and vector nonet meson fields~\cite{Cao:2023gfv}:
\begin{equation}
  \begin{aligned}
    \mathcal{P}=
    \mqty(
      \frac {\pi^0} {\sqrt {2}} + \frac {\eta} {\sqrt {6}} & \pi^+ & K^+ \\
      \pi^- & -\frac {\pi^0} {\sqrt {2}} + \frac {\eta} {\sqrt {6}} & K^0 \\
      K^-& {\bar K}^0 & -\sqrt {\frac{2}{3}} \eta \\
    ),\quad
    \mathcal{V}=\mqty(\frac{\rho^0} {\sqrt {2}}+\frac {\omega} {\sqrt {2}}&\rho^+ & K^{*+} \\
      \rho^- & -\frac {\rho^0} {\sqrt {2}} + \frac {\omega} {\sqrt {2}} & K^{*0} \\
      K^{*-}& {\bar K}^{*0} & \phi \\
    ).  \\
  \end{aligned}
\end{equation}

\begin{figure}[!ht]
  \centering
  \subfigure[]{\includegraphics[width=0.6\textwidth]{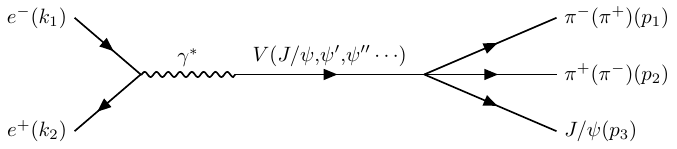}}\\
  \subfigure[]{\includegraphics[width=0.6\textwidth]{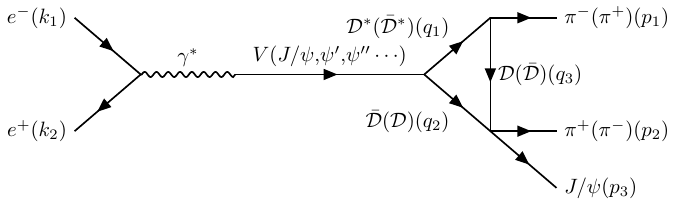}}\\
  \subfigure[]{\includegraphics[width=0.6\textwidth]{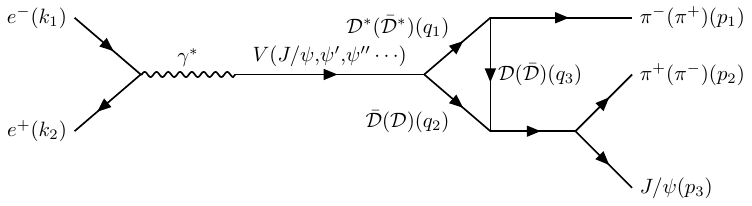}}
  \caption{Schematic diagrams for the $e^{+}e^{-}\to J/\psi\,\pi^{+}\pi^{-}$ process, where $\mathcal{D}=(D^{0},D^{+})$ and $\bar{\mathcal{D}}=(\bar{D}^{0},D^{-})$. (a) shows the tree-level diagram, (b) corresponds to the loop diagram without intermediate resonance in $\mathcal{D} \bar{\mathcal{D}} \to J / \psi \pi $, and (c) represents the loop diagram with an intermediate resonance in $\mathcal{D} \bar{\mathcal{D}} \to J / \psi \pi$.}\label{fig:jpsipipi}
\end{figure}

The process $e^{+}e^{-}\to J/\psi\,\pi^{+}\pi^{-}$ is illustrated in \cref{fig:jpsipipi}. We focus on the energy region $\sqrt{s}=3.8\sim 4.0\,\mathrm{GeV}$, where possible enhancements due to triangle singularities may occur. Both tree-level and loop contributions are considered. For the loop diagrams, two scenarios are taken into account: (1) direct coupling of $\mathcal{D}\bar{\mathcal{D}}$ to $J/\psi\,\pi$; (2) $\mathcal{D}\bar{\mathcal{D}}$ coupling to $J/\psi\,\pi$ via an intermediate state.

First, we consider the tree-level contribution, where the intermediate charmonium state directly couples to $J/\psi\,\pi\pi$ via a contact vertex. The amplitude can be written as
\begin{equation}
  i \mathcal{M}_{\text{tree}}
  =g_{\text{tree}}\bar{v}(k_2)(ie)\gamma^{\mu}u(k_1)\frac{g_{\mu\nu}}{s}\sum_{V} \frac{e m^{2}_{V}}{f_{V}} \frac{\varepsilon_{3}^{\nu}}{s-m^{2}_{V}+i m_{V}\Gamma_{V}}
\end{equation}
where $\sum_{V}$ denotes the sum over all possible intermediate vector charmonium states in the framework of vector meson dominance (VMD) model, and $f_V$ is the vector meson decay constant to be determined by $V\to e^+e^-$; $\varepsilon_{3}$ is the polarization vector of $J/\psi$.

For the coupling between the final state $\mathcal{D} \bar{\mathcal{D}}$ and $J/\psi\,\pi^{\pm}$, if no intermediate resonance is considered, the amplitude can be described by a contact vertex as
\begin{equation}
  i \mathcal{M}(\mathcal{D} \bar{\mathcal{D}} \to J/\psi\,\pi^{+} )= i g_{c}\varepsilon_{\mu \alpha \beta \sigma}\varepsilon_{3}^{\beta}(q_{2}-q_{3})^{\mu}p_{2}^{\alpha}p_{3}^{\sigma},
\end{equation}
where $g_{c}$ is the effective coupling constant. If an intermediate vector resonance is included, the amplitude takes the form
\begin{equation}
  i \mathcal{M}(\mathcal{D} \bar{\mathcal{D}}\to J/\psi\,\pi^{+}) = i g_{r}(q_2-q_{3})_{\mu}\frac{g^{\mu\nu}}{p^{2}-m_{V'}^{2}+i m_{V'}\Gamma_{V'}} \varepsilon_{\nu \alpha \beta \sigma}\varepsilon_{3}^{\beta}p_{2}^{\alpha}p_{3}^{\sigma},
\end{equation}
where $m_{V'}$ and $\Gamma_{V'}$ are the mass and width of the intermediate resonance, respectively, and $g_{r}$ is the product of the coupling constants at the $\mathcal{D} \bar{\mathcal{D}} V'$ and $J/\psi\,\pi V'$ vertices.

Taking into account both tree-level and loop contributions, the total amplitude for $e^{+}e^{-}\to J/\psi\,\pi^{+}\pi^{-}$ can be written as
\begin{equation}
  i\mathcal{M}^{(\text{C/R})}= \bar{v}(k_2)ie\gamma^{\mu}u(k_1)\frac{ g_{\mu\nu}}{s}\sum_{V} \frac{e m^{2}_{V}}{f_{V}}\frac{g^{\nu \alpha}}{s-m^{2}_{V}+i m_{V}\Gamma_{V}}\qty(\sum_{j}\mathcal{M}_{j\alpha}^{(\text{C/R})}+g_{\text{tree}}\varepsilon^{\psi}_{\alpha}).
\end{equation}
Here, the superscript C denotes the loop contribution where $\mathcal{D} \bar{\mathcal{D}}$ directly couples to $J/\psi\,\pi$, while R represents the loop contribution mediated by an intermediate resonance in the $\mathcal{D} \bar{\mathcal{D}} \to J/\psi\,\pi$ transition. The loop amplitude for the case where $\mathcal{D} \bar{\mathcal{D}}$ directly couples to $J/\psi\,\pi$ is given by
\begin{equation}
  i \mathcal{M}_{j \alpha}^{\text{C}}
  =\int \frac{d^{4}q_{3}}{(2\pi)^{4}}\frac{ g_{c}g_{\psi \mathcal{D}^{*} \mathcal{D}}g_{\mathcal{D}^{*} \mathcal{D}\mathcal{P}}\varepsilon_{\alpha \rho \sigma \lambda }p^{\rho}(q_2-q_{1})^{\sigma}\qty(g^{\lambda\kappa}-\frac{q_{1}^{\lambda}q_{1}^{\kappa}}{m_1^{2}})p_{1\kappa}\varepsilon_{\mu \delta\beta \tau}\varepsilon_{\psi}^{\beta}(q_{2}-q_{3})^{\mu}p_{2}^{\delta}p_{3}^{\tau}}{(q_1^{2}-m_1^{2})(q_2^{2}-m_2^{2})(q_3^{2}-m_3^{2})}\mathcal{F}(q_{i}^{2}) \ .
\end{equation}
The loop amplitude for the case where $\mathcal{D} \bar{\mathcal{D}}$ couples to $J/\psi\,\pi$ via an intermediate resonance is
\begin{equation}
  \mathcal{M}_{j \alpha}^{\text{R}}= \frac{\mathcal{M}_{j \alpha}^{\text{C}}}{(p_2+p_3)^{2}-m^{2}_{V'}+i m_{V'}\Gamma_{V'}}\cdot \frac{g_{r}}{g_{c}} .
\end{equation}
In the above equations $\mathcal{F}(q_{i}^{2})$ is a form factor adopted for cutting off the ultraviolet divergence in the loop integrals, and has the following form
\begin{equation}\label{formfactor}
  \mathcal{F}(q_{i}^{2})=\prod_{i}\qty(\frac{\Lambda_{i}^{2}-m_i^{2}}{\Lambda_{i}^{2}-q_{i}^{2}}),
\end{equation}
where $\Lambda_{i}\equiv m_{i}+\alpha \Lambda_{\text{QCD}}$ with the $m_i$ the mass of the $i$-th internal particle, and the QCD energy scale $\Lambda_{\text{QCD}}=220\, \mathrm{MeV}$ with $\alpha=1\sim 2$ as the cutoff parameter~\cite{Guo:2010ak,Cao:2023csx,Wang:2025fzj,Cao:2023gfv}.

The above loop integrals can be expressed in terms of Passarino-Veltman coefficient functions~\cite{Passarino:1978jh}. The loop amplitude can be formally written as
\begin{equation}
  \mathcal{M}_{j }^{\alpha}=C_{1} \varepsilon^{\alpha}_{3}+C_{2}p_{2}^{\alpha} (\varepsilon_{3}\cdot p_1)+C_{3}p_{2}^{\alpha}(\varepsilon_{3}\cdot p_{2})+C_{4}p_{3}^{\alpha} (\varepsilon_{3}\cdot p_{1})+C_{5}p_{3}^{\alpha}(\varepsilon_{3}\cdot p_{2})\,,
\end{equation}
where $C_{i}\ (i=1,2,3,4,5)$ are combinations of Passarino-Veltman coefficient functions. The numerical evaluation of the loop integrals is performed using LoopTools~\cite{Hahn:1998yk}.

\subsection{Relevant Coupling Constants}
The factor $e/f_{V}$ is determined by the VMD model
\begin{equation}
  \frac{e}{f_{V}}=\sqrt{\frac{3\Gamma_{V\to e^{+}e^{-}}}{2 \alpha_e |\mathbf{p}_{e}|}}\,,
\end{equation}
where $\alpha_e$ is the fine-structure constant, $|\mathbf{p}_{e}|$ is the momentum of the electron in the rest frame of the initial vector meson, and $\Gamma_{V\to e^{+}e^{-}}$ is the partial width for the vector meson decay into $e^{+}e^{-}$.
The coupling $g_{\mathcal{D}^{*}\mathcal{D}\pi}$ is determined by heavy quark symmetry~\cite{Wang:2011yh,Cheng:2004ru,Casalbuoni:1996pg,Wang:2025fzj}:
\begin{equation}
  \frac{g_{D^{*}D\pi}}{\sqrt{m_{D}m_{D^{*}}}}=\frac{2g}{f_{\pi}},
\end{equation}
where $g=0.59$ and $f_{\pi}=132\, \mathrm{MeV}$ are adopted.
For the coupling constants between charmonium and charmed meson pairs, we employ heavy quark symmetry~\cite{Casalbuoni:1996pg,Cheng:2004ru,Wang:2011yh}:
\begin{equation}
  g_{\psi \mathcal{D}^{*} \bar{D}}=\frac{g_{\psi \mathcal{D} \bar{\mathcal{D}}}}{\tilde{M}_{\mathcal{D}}},\quad \tilde{M}_{\mathcal{D}}=\sqrt{ m_{\mathcal{D}}m_{\mathcal{D}^{*}}}
\end{equation}
Since $J/\psi$ and $\psi(3686)$ are below the $D\bar{D}$ threshold, their couplings are extracted from the heavy quark symmetry~\cite{Wang:2011yh,Zhang:2009kr}. The couplings for other charmonium states are determined from experimental data of $e^{+}e^{-}\to D\bar{D}$~\cite{Zhang:2009gy,ParticleDataGroup:2024cfk}.
\section{Numerical Results and Discussions}\label{sec:results}
Based on the theoretical framework described above, we fit the lineshape of the total cross section for $e^{+}e^{-}\to J/\psi\,\pi^{+}\pi^{-}$ to extract the relevant parameters and subsequently compute the $J/\psi\pi$ invariant mass spectrum. This will allow us to identify observable effects from either kinematic singularity or genuine resonance.

First, we list the parameters used in the fit, which are shown in \cref{tab:1}.  For the case where $\mathcal{D} \bar{\mathcal{D}} \to J / \psi \pi$ involves an intermediate resonance, we adopt the resonance parameters as $m_{V'}=2m_{D}=3.728~\mathrm{GeV},~\Gamma_{V'}=20~\mathrm{MeV}$. This is an analogue of $G(3900)$ for the $D\bar{D}$ system, where the $P$-wave interaction between $D$ and $\bar{D}$ may form a resonance with $I=1$. Such an assumption will only have a meaning of illustration since so far theoretical studies of the $P$-wave $D\bar{D}$ scatterings and hidden charm tetraquarks are strongly model-dependent. We will show that any signal arising from such a dynamical scenario can manifest itself in the $J/\psi\pi$ invariant mass spectrum.

\begin{table}[!ht]
  \centering\caption{The relevant parameters for the intermediate charmonium states.}\label{tab:1}
  \begin{ruledtabular}
    \begin{tabular}{cccccc}
      Particle&$J / \psi$&$\psi(3686)$&$\psi(3770)$&$\psi(4040)$&$\psi(4160)$\\\hline
      Mass($\mathrm{MeV}$)&$3096.9$&$3686.1$&$3773.7$&$4040$&$4191$\\
      $\Gamma(\mathrm{MeV})$&$0.0926$&$0.286$&$27.2$&$84$&$69$\\
      $\mathrm{BR}(\psi\to e^{+}e^{-})$&$5.971\times 10^{-2}$&$7.94\times 10^{-3}$&$9.6\times 10^{-6}$&$1.02\times 10^{-5}$&$6.9\times 10^{-6}$\\
      $e/ f_{\psi}$&$2.71\times 10^{-2}$&$1.59\times 10^{-2}$&$5.33\times 10^{-3}$&$9.34\times 10^{-3}$&$6.83\times 10^{-3}$\\
      $g_{\psi \mathcal{D} \bar{\mathcal{D}}^{*}}(\mathrm{GeV}^{-1})$&$3.8$&$4.3$&$6.2$&$1.6$&$0.71$
    \end{tabular}
  \end{ruledtabular}
\end{table}

To proceed, we will fit the lineshape of the total cross section with $\alpha =1.5$ as a typical cutoff parameter. We then consider two scenarios. In scenario-I the transition, $\mathcal{D}\bar{\mathcal{D}}\to J/\psi\pi$, is via a contact interaction, and in scenario-II, the $\mathcal{D}\bar{\mathcal{D}}\to J/\psi\pi$ transition is via an intermediate resonance.

For each scenario we will employ two fitting schemes. For the first scheme (Fit-1) we will include only the intermediate charmonium states listed in \cref{tab:1}. For the second scheme (Fit-2) we will include, in addition to the states in \cref{tab:1}, an extra intermediate state with mass $3905\ \mathrm{MeV}$ and width $346\ \mathrm{MeV}$ as obtained in the experimental fit.

As follows, we first consider scenario-I and investigate the fittings of these two schemes. In the first scheme (Fit-1), i.e. by including only the intermediate charmonium states listed in \cref{tab:1}, it shows that the model cannot reproduce the subset of experimental points used in the experimental fits (i.e. the circle green points in \cref{fig:2})~\cite{BESIII:2022qal,BESIII:2016bnd}. The fitting result is illustrated by the blue solid curve in \cref{fig:2}. In the energy interval 3.85$-$3.95$\ \mathrm{GeV}$ the fit tends to follow the data point that was not included in the original experimental fit (i.e. the triangle red point in \cref{fig:2}).

In the second scheme (Fit-2) we include, in addition to the states in \cref{tab:1}, an extra intermediate state with mass $3905\ \mathrm{MeV}$ and width $346\ \mathrm{MeV}$ as obtained in the experimental fit. To limit the number of free parameters, only its VMD coupling $e/f_{V}$ is treated as a free parameter, while its coupling to the $\mathcal{D}^{*}\mathcal{D}$ channel is fixed to
$g_{V\mathcal{D}^{*}\mathcal{D}}=1\ \mathrm{GeV}^{-1}$.
In principle, interference among different resonances should be included. However, owing to the limited experimental data we introduce only a single relative phase parameter $\theta$ between the two resonances that are closest to the energy region considered, $\psi(3770)$ and $\psi(4040)$, in the fit. The fitting result is presented by the brown dashed curve in \cref{fig:2}.

\subsection{Fit results}
\begin{figure}[!ht]
  \centering
  \subfigure[]{\includegraphics[width=0.497\textwidth]{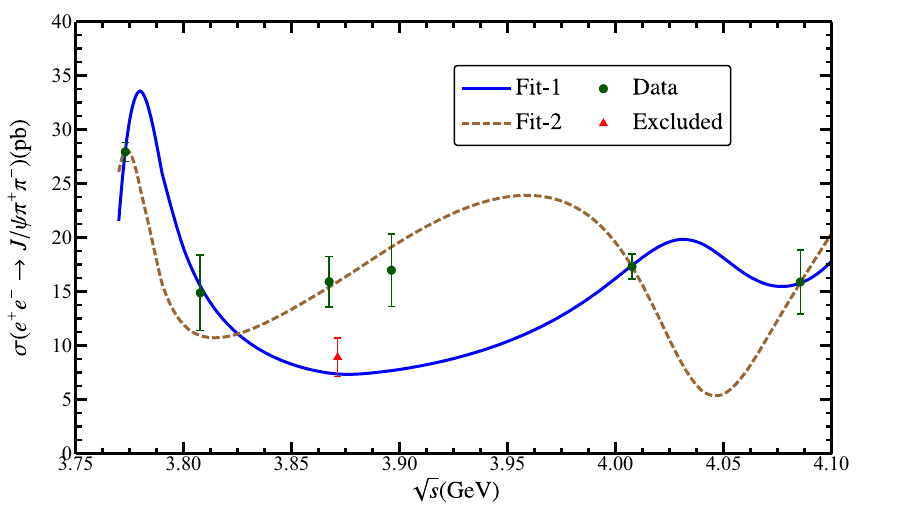}\label{fig:2_a}}
  \subfigure[]{\includegraphics[width=0.497\textwidth]{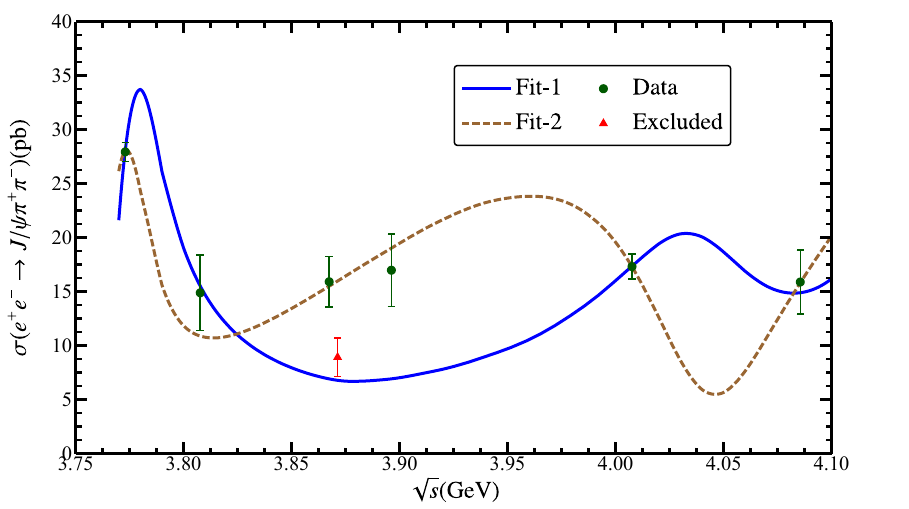}\label{fig:2_b}}
  \caption{Fitted lineshape of the total cross section for $e^{+}e^{-}\to J/\psi\,\pi^{+}\pi^{-}$. Panel (a) illustrates the results for scenario-I with $\mathcal{D}\bar{\mathcal{D}}\to J/\psi\,\pi$ by a contact interaction. Panel (b) illustrates the results for scenario-II with $\mathcal{D}\bar{\mathcal{D}}\to J/\psi\,\pi$ via an intermediate resonance.}\label{fig:2}
\end{figure}

The fitted parameters for the cases where the $\mathcal{D}\bar{\mathcal{D}}\to J/\psi\pi$ transition is described by a contact term and via an intermediate resonance are listed in \cref{tab:2,tab:3}, respectively. The corresponding fitted lineshapes are shown in \cref{fig:2}.
\begin{table}[!ht]
  \centering\caption{The results of fitting parameters in scenario-I with a contact vertex for $\mathcal{D} \bar{\mathcal{D}} \to J / \psi \pi $.}\label{tab:2}
  \begin{ruledtabular}
    \begin{tabular}{cccccc}
      Parameters&$g_{\text{tree}}$&$g_{c}(\mathrm{GeV}^{-3})$&$\theta(^\circ)$&$e/f_{V}$&$\chi^{2}/\text{d.o.f}$\\\hline
      Fit-1&$5.94\pm 0.11$&$18.3\pm 3.7$&$-125\pm 11$&$\setminus$&$0.4$\\
      Fit-2&$3.83\pm 0.21$&$10.7\pm 2.9$&$-142\pm 5$&$(7.38\pm 0.89)\times 10^{-2}$&$0.9$\\
    \end{tabular}
  \end{ruledtabular}
\end{table}
\begin{table}[!ht]
  \centering\caption{The results of fitting parameters in scenario-II with an intermediate resonance in $\mathcal{D} \bar{\mathcal{D}} \to J / \psi \pi $.}\label{tab:3}
  \begin{ruledtabular}
    \begin{tabular}{cccccc}
      Parameters&$g_{\text{tree}}$&$g_{r}(\mathrm{GeV}^{-1})$&$\theta(^\circ)$&$e/f_{V}$&$\chi^{2}/\text{d.o.f}$\\\hline
      Fit-1&$6.04\pm 0.14$&$4.11\pm 0.94$&$-130\pm 14$&$\setminus$&$0.8$\\
      Fit-2&$3.85\pm 0.21$&$2.86\pm 0.71$&$-141\pm 5$&$(7.46\pm 0.84)\times 10^{-2}$&$0.9$\\
    \end{tabular}
  \end{ruledtabular}
\end{table}

As shown by the brown dashed curve in \cref{fig:2}, the presence of an intermediate state coupling to the $\mathcal{D}^{*}\mathcal{D}$ channel near $\sqrt{s}\simeq 3.9~\mathrm{GeV}$ can have a sizable effect on the lineshape of the total cross section, although it still cannot accommodate the triangle red data point. Comparing these two fitting schemes, the effects caused by the inclusion of an additional vector charmonium state around $3.9~\mathrm{GeV}$ are rather significant. It can strongly enhance the cross section to better accommodate the two data points around 3.9 GeV (circle green points). Also, different interfering structures can arise in the energy region between $4.0\sim 4.1$ GeV.
However, due to the limited experimental data for $e^{+}e^{-}\to J/\psi\,\pi^{+}\pi^{-}$ in the vicinity of $3.9~\mathrm{GeV}$, it is not possible to unambiguously determine whether a resonance should be present here. If such a state does exist, further investigations are needed to disentangle its properties. Namely, whether it corresponds to a pole structure due to the $P$-wave $D\bar{D}^*+c.c.$ interaction~\cite{Lin:2024qcq}, or just an open channel effect~\cite{Zhang:2010zv,Zhang:2009gy}. More comprehensive and precise measurements in this energy region are also needed to clarify the underlying dynamics.

\begin{figure}[!ht]
  \centering
  \subfigure[]{\includegraphics[width=0.497\textwidth]{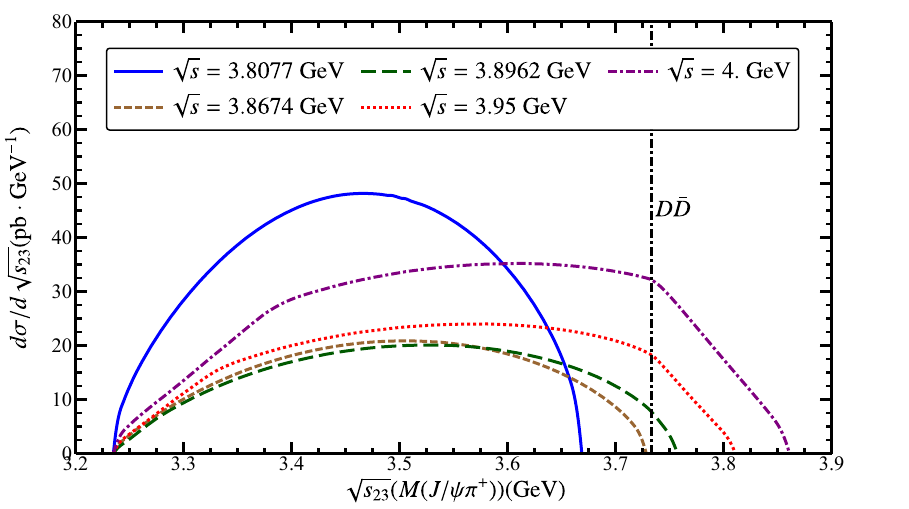}\label{fig:3_a}}
  \subfigure[]{\includegraphics[width=0.497\textwidth]{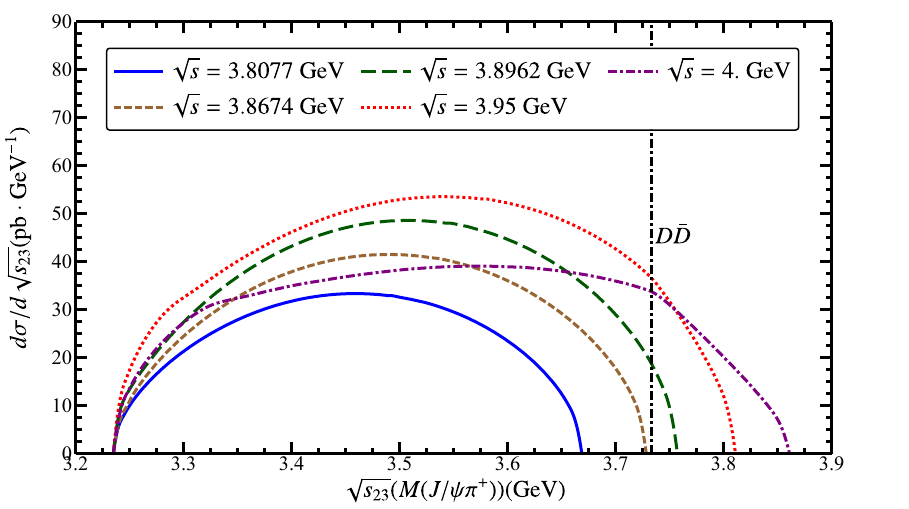}\label{fig:3_b}}
  \caption{Invariant mass spectrum of $J / \psi \pi^{+}$ (identical for $J / \psi \pi^{-}$) where $\mathcal{D} \bar{\mathcal{D}} \to J / \psi \pi$ by a contact interaction for different $\sqrt{s}$. Panel (a) corresponds to the results of Fit-1. Panel (b) corresponds to the results of Fit-2.}\label{fig:3}
\end{figure}

\begin{figure}[!ht]
  \centering
  \subfigure[]{\includegraphics[width=0.497\textwidth]{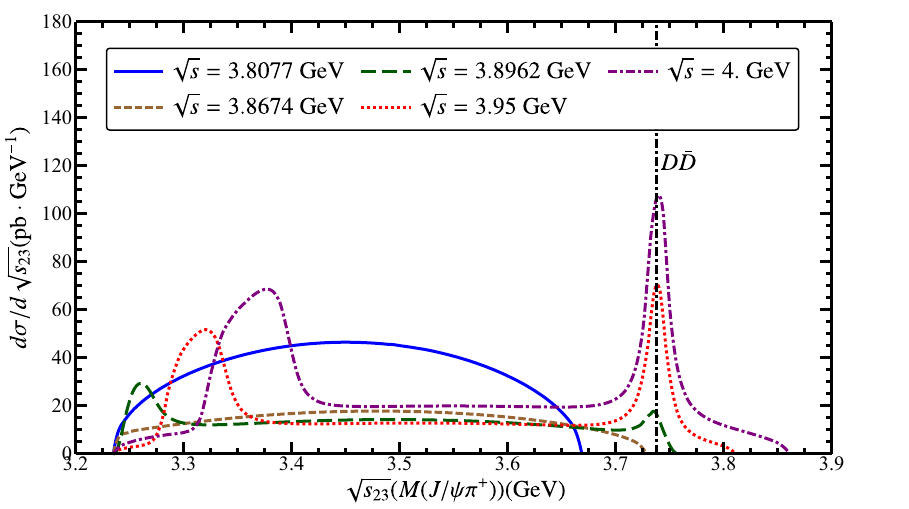}\label{fig:4_a}}
  \subfigure[]{\includegraphics[width=0.497\textwidth]{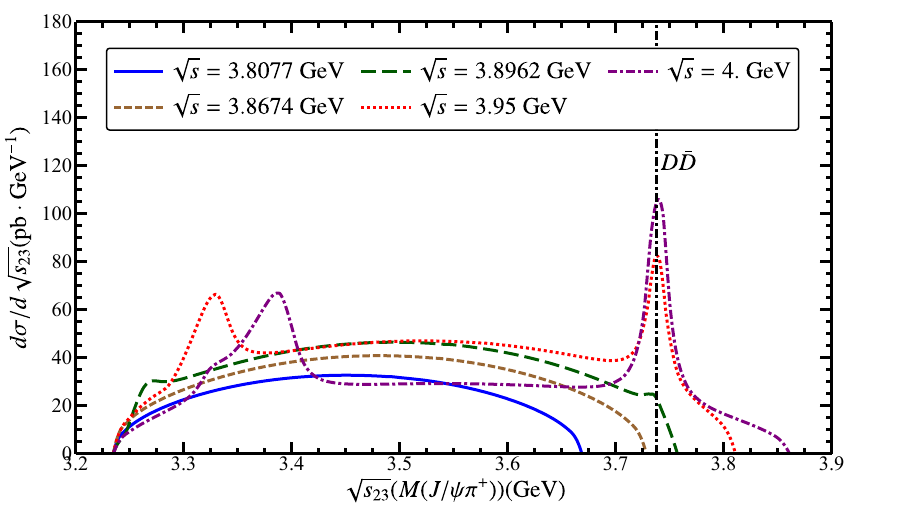}\label{fig:4_b}}
  \caption{Invariant mass spectrum of $J / \psi \pi^{+}$ (identical for $J / \psi \pi^{-}$) where $\mathcal{D} \bar{\mathcal{D}} \to J / \psi \pi$ via an intermediate resonance for different $\sqrt{s}$. Panel (a) corresponds to the results of Fit-1. Panel (b) corresponds to the results of Fit-2.}\label{fig:4}
\end{figure}

We also investigate the effects arising from the loop transitions. As mentioned earlier, we consider two scenarios in $\mathcal{D} \bar{\mathcal{D}}\to J / \psi \pi$, i.e. either the transition is via a contact interaction, or via an intermediate vector state. The results for scenario-I have been presented in \cref{fig:2_a}. The results for scenario-II are presented in \cref{fig:2_b} as a comparison. Interestingly, it shows that the fitting parameters and lineshapes are nearly identical, regardless of whether the $\mathcal{D}\bar{\mathcal{D}}\to J/\psi\,\pi$ transition is described by a contact interaction or mediated by an intermediate resonance. This indicates that the total cross section alone cannot distinguish between these two mechanisms. Therefore, we proceed to examine the invariant mass spectrum of the $J/\psi\,\pi$ final state, as shown in \cref{fig:3,fig:4}.


For the case of scenario-I where $\mathcal{D} \bar{\mathcal{D}} \to J/\psi \pi$ is described by a contact interaction, the $J/\psi \pi^+$ invariant mass spectra calculated in Fit-1 at different c.m. energies are shown in \cref{fig:3_a}. It shows that when the c.m. energy is sufficiently large such that $M(J / \psi \pi^{+})$ reaches the $D \bar{D}$ threshold, the lineshape of the invariant mass spectrum starts to deviate from the phase space and a CUSP structure appears at the $D\bar{D}$ threshold for $\sqrt{s}>3.95$ GeV. This is attributed to the enhancement caused by the TS~\cite{Landau:1959fi,Cutkosky:1960sp,Coleman:1965xm,Wu:2011yx,Wang:2013cya,Wang:2013hga,Liu:2017rhx,Du:2019idk}. It should be noted that the exclusive transition amplitude of the triangle loop will produce a narrow peak near the $D\bar{D}$ threshold due to the TS mechanism. However, since the loop contribution is much smaller than that from the tree amplitudes, it can only produce a small cusp effect in the invariant mass spectra.

In \cref{fig:3_b} the Fit-2 result for the $J/\psi\pi^+$ spectra in scenario-I is presented to compare with that of Fit-1. It shows that the inclusion of the charmonium vector can cause quite significant changes to the absolute magnitudes of the $J/\psi\pi^+$ spectra. However, the behavior is very similar to the phase space. The effects from the TS also turn out to be negligible because of the much smaller contribution from the triangle loops than from the tree-diagram contribution.

For scenario-II where $\mathcal{D} \bar{\mathcal{D}} \to J / \psi \pi$ proceeds via an intermediate resonance, it can be seen in \cref{fig:4} that when the c.m. energy is sufficiently large such that $M(J / \psi \pi^{+})$ reaches the $D \bar{D}$ threshold, the lineshape of the two-body spectrum will deviate from the phase space significantly, and a pronounced peak near the $D\bar{D}$ threshold will be produced for both Fit-1 and Fit-2. This is attributed to the pole structure introduced to the $\mathcal{D} \bar{\mathcal{D}} \to J / \psi \pi$ transition. The kinematic reflection effects can also be seen in the lower energy region. As shown by \cref{fig:4} the $J/\psi\pi^+$ invariant mass spectrum turns to be sensitive to the presence of the pole structure which can couple to the $D\bar{D}$ channel, it suggests that the $J/\psi\pi^+$ invariant mass spectrum can be an ideal observable for the search of the exotic isospin-1 charmonia with $(I,J^{P(C)})=(1,1^{-(-)})$.

\section{Summary}\label{sec:summary}
In this work, we investigate the total-cross-section lineshape of $e^{+}e^{-}\to J/\psi\,\pi^{+}\pi^{-}$ and the two-body invariant-mass spectrum of $J/\psi\,\pi$ in the energy region $\sqrt{s}=3.8\sim 4.0\,\mathrm{GeV}$. This is the energy region where the $G(3900)$ structure is located. Within a unified framework to take into account the $D^*\bar{D}+c.c.$ open channel effects via rescatterings, we investigate the relative strengths between the tree and triangle loop amplitudes by fitting to experimental data. We consider two mechanisms for $\mathcal{D}\bar{\mathcal{D}}\to J/\psi\pi$. One is through a contact interaction, and the other is through an intermediate resonance. It shows that the TS mechanism can produce some effects even for a contact interaction for $\mathcal{D}\bar{\mathcal{D}}\to J/\psi\pi$. Meanwhile, if $\mathcal{D}\bar{\mathcal{D}}\to J/\psi\pi$ occurs via an intermediate resonance, its signal in the $J/\psi\pi$ invariant mass spectrum will be strongly enhanced by the TS.
We show that the analysis of the $J/\psi\,\pi$ invariant-mass distribution can provide a practical way to distinguish whether the observed structure originates from the TS mechanism or from a genuine intermediate charmonium resonance with exotic quantum numbers of $(I,J^{P(C)})=(1,1^{-(-)})$. This test is logically distinct from the role assigned to $G(3900)$ in the production amplitude: $G(3900)$ characterizes the overall $e^+e^-\to D\bar D$ lineshape, whereas the pole tested here belongs to the $D\bar D\to J/\psi\pi$ transition and would be manifested in the $J/\psi\pi$ invariant-mass distribution. 
In the present fits, the total cross section alone does not sharply separate the contact and pole scenarios, while their predicted $J/\psi\pi$ spectra are markedly different. These results are important for understanding the dynamical mechanism of $e^{+}e^{-}\to J/\psi\,\pi^{+}\pi^{-}$ and the nature of the relevant resonant states. More precise experimental measurements in the future will be crucial for further testing our conclusions and clarifying the underlying physics.

\acknowledgments
Useful discussion with Chang-Zheng Yuan is acknowledged. This work is supported by the National Natural Science Foundation of China under Grant No. 12235018.

\bibliography{0.Jpsi_pi_pi.bib}
\end{document}